\documentclass[12pt]{article} 
\usepackage[reqno]{amsmath} 
\usepackage{bbm} 
\usepackage{epsfig} 
\usepackage{array} 
\usepackage{float} 
\usepackage{amstext} 
 
\usepackage{a4} 

\usepackage{a4wide} 
\usepackage{wasysym} 
\def\be{\begin{equation}} 
\def\ee{\end{equation}} 
\def\gs{\mathrel{ 
   \rlap{\raise 0.511ex \hbox{$>$}}{\lower 0.511ex \hbox{$\sim$}}}} 
\def\ls{\mathrel{ 
   \rlap{\raise 0.511ex \hbox{$<$}}{\lower 0.511ex \hbox{$\sim$}}}} 
 
\newcommand{\onbb}{neutrinoless double beta decay } 
\newcommand{\ba}{\begin{array}{c}} 
\newcommand{\baz}{\begin{array}{cc}} 
\newcommand{\bad}{\begin{array}{ccc}} 
\newcommand{\bea}{\begin{equation} \begin{array}{c}} 
\newcommand{\eea}{ \end{array} \end{equation}} 
\newcommand{\ea}{\end{array}} 
\newcommand{\D}{\displaystyle} 
\newcommand{\dms}{\mbox{$\Delta m^2_{\odot}$}} 
\newcommand{\dma}{\mbox{$\Delta m^2_{\rm A}$}} 
\newcommand{\meff}{\mbox{$\langle m \rangle$}}

\def\gsim{\ \rlap{\raise 2pt\hbox{$>$}}{\lower 2pt \hbox{$\sim$}}\ } 
\def\lsim{\ \rlap{\raise 2pt\hbox{$<$}}{\lower 2pt \hbox{$\sim$}}\ }

\begin{document} 
 
\title{\vspace{-2cm} 
%\hfill {\small hep-ph/0612328} 
%\hfill{\small \today} 
\vskip 0.4cm 
\bf \Large 
Corrections to Tri-bimaximal Neutrino Mixing: Renormalization  
and Planck Scale Effects 
} 
\author{ 
Amol~Dighe$^a$\thanks{email: \tt amol@theory.tifr.res.in}~~,~~ 
Srubabati~Goswami$^b$\thanks{email: \tt sruba@mri.ernet.in}~~,~~ 
Werner~Rodejohann$^c$\thanks{email: \tt werner.rodejohann@mpi-hd.mpg.de} 
\\\\ 
{\normalsize \it $^a$Tata Institute of Fundamental Research,}\\ 
{\normalsize \it Homi Bhabha Road, Mumbai 400005, India}\\ \\ 
{\normalsize \it $^b$Harish--Chandra Research Institute, Chhatnag Road,}\\ 
{\normalsize \it Jhunsi, Allahabad 211 019, India}\\ \\ 
{\normalsize \it $^c$Max--Planck--Institut f\"ur Kernphysik,}\\ 
{\normalsize \it  Postfach 103980, D--69029 Heidelberg, Germany} 
} 
\date{} 
\maketitle 
\thispagestyle{empty} 
\vspace{-0.8cm} 
\begin{abstract} 
%\noindent 
 
We study corrections to tri-bimaximal (TBM) neutrino mixing from  
renormalization group (RG) running and from Planck scale effects.  
We show that while the RG effects are negligible in the  
standard model (SM), for quasi-degenerate neutrinos  
and large $\tan\beta$ in the minimal supersymmetric 
standard model (MSSM) all three mixing angles may  
change significantly.  
In both these cases, the direction of the modification of  
$\theta_{12}$ is fixed, 
while that of $\theta_{23}$ is determined by the  
neutrino mass ordering.  
The Planck scale effects can also change $\theta_{12}$ 
up to a few degrees in either direction  
for quasi-degenerate neutrinos. 
These effects may dominate over the RG effects in the SM,  
and in the MSSM with small $\tan \beta$.  
The usual constraints on neutrino masses, Majorana phases  
or $\tan \beta$ stemming from RG running arguments  
can then be relaxed.
We quantify the extent of Planck effects on the mixing angles 
in terms of ``mismatch phases'' which break the symmetries
leading to TBM. In particular, we show that
when the mismatch phases vanish, the mixing angles 
are not affected in spite of the Planck scale contribution.
Similar statements may be made for 
$\mu$--$\tau$ symmetric mass matrices.

%If we include only the Majorana phases in the neutrino mixing matrix,  
%Planck scale effects for the mixing angles in TBM are absent.  
%Similar statements can be made for 
%Sizable Planck effects require the so-called  
%``unphysical phases'' of the leptonic mixing matrix to be  
%non-vanishing, which underlines the importance of these phases.  

\end{abstract} 
 
%\preprint{TIFR/TH/06-39} 
 
\newpage

\section{\label{sec:intro}Introduction} 
 
Neutrino mixing \cite{reviews} is a consequence of a non-trivial structure  
of the neutrino mass matrix. This mass matrix is generated  
by the following dimension five operator:  
\be \label{eq:5} 
{\cal L}_5 = \frac{\mu_{\alpha \beta}}{\Lambda} \,  
\left(L_\alpha \, \Phi \right) \, \left(L_\beta \, \Phi \right)  
+ h.c.  
\ee 
Here $\mu_{\alpha \beta}$ is a coupling matrix,  
$\Lambda$ some high energy scale,  
$L_\alpha$ is the lepton doublet with $\alpha \in \{ e, \mu, \tau \}$, 
and $\Phi$ is the Higgs doublet (in the MSSM, $\Phi$ is the 
Higgs doublet that couples to the up-type fermions).  
After the electroweak symmetry breaking  
($\langle \Phi\rangle \equiv v/\sqrt{2} \simeq 174$ GeV for the SM, 
$\langle \Phi \rangle \equiv v \sin \beta / \sqrt{2}$ for the MSSM), 
Eq.~(\ref{eq:5}) gives rise to the neutrino mass matrix  
\be \label{eq:mnu} 
(m_\nu)_{\alpha \beta} = \mu_{\alpha \beta} \,  
\frac{\langle \Phi \rangle^2}{\Lambda}   
= \left(U^* \, m_\nu^{\rm diag} \, U^\dagger\right)_{\alpha \beta}~, 
\ee 
where $U$ is the leptonic mixing, or  
Pontecorvo-Maki-Nakagawa-Sakata (PMNS), matrix  
in the basis in which the charged lepton mass matrix is real and diagonal.  
The neutrino masses are contained in  
$m_\nu^{\rm diag} = {\rm diag}(m_1, m_2, m_3)$. 
The value of $\Lambda$ may be taken to be the  
seesaw scale, $\Lambda \sim 10^{12}$ GeV,  
where the underlying flavor structure 
of the theory is implemented. 
However, measurements take place at low scale, therefore the predictions of  
any neutrino mass model have to be evolved down to low energy through  
renormalization group (RG) running  
\cite{rgpapers, chankowski}. 
As the large majority of models is generated at a high energy, 
RG effects are a generic feature.

Another guaranteed correction to  
the dimension five operator in  
Eq.~(\ref{eq:5}) is an additional operator of the  
same form but with $\Lambda$ identified as  
the Planck mass $M_{\rm Pl} = 1.2 \cdot 10^{19}$ GeV.  
Since gravity does not distinguish between flavors,
the operator is expected to be flavor democratic. 
The presence of such a term in the Lagrangian  
gives rise to additional contributions to the neutrino mass matrix after  
the electroweak symmetry is broken.  
While too small to  
be responsible for the leading structures of the neutrino mass matrix, those  
Planck scale effects can have observable consequences  
as well \cite{planck,ue3,new}. In the present paper we  
perform a comparative study  
of  both  
renormalization and Planck scale effects. 
 
For this analysis we will choose one particular and very interesting neutrino  
mixing scheme which is  
compatible with all data, the tri-bimaximal mixing (TBM),  
defined by \cite{tbm}  
\be \label{eq:Utri}  
U = {\rm diag}(e^{i \phi_1}, e^{i \phi_2}, e^{i \phi_3}) \, \left( 
\bad  
\sqrt{\frac{2}{3}} & \sqrt{\frac{1}{3}} & 0 \\[0.2cm] 
-\sqrt{\frac{1}{6}} & \sqrt{\frac{1}{3}} & -\sqrt{\frac{1}{2}}  \\[0.2cm] 
-\sqrt{\frac{1}{6}} & \sqrt{\frac{1}{3}} & \sqrt{\frac{1}{2}}   
\ea  
\right)  
\, {\rm diag}(e^{-i \alpha_1/2}, e^{-i \alpha_2/2}, e^{-i \alpha_3/2})~.  
\ee 
Here  
$P \equiv  
{\rm diag}(e^{-i \alpha_1/2}, e^{-i \alpha_2/2}, e^{-i \alpha_3/2})$  
contains the Majorana phases, one of which can be rephased away. 
We further have an additional phase matrix  
$Q \equiv {\rm diag}(e^{i \phi_1}, e^{i \phi_2}, e^{i \phi_3})$, 
which is usually phased away by a redefinition of the charged lepton 
fields in order to bring $U$ in the standard form.
Due to this, the entries in $Q$ have received the (possibly unfair) 
title ``unphysical phases''.  
However a complete theory of neutrino masses, especially like the
one we are considering here which has more than one independent 
operator contributing to the neutrino mass, 
must be able to predict these phases.
In the basis in which the flavor democratic Planck scale
contribution to the neutrino mass matrix is completely real,
the phases $\phi_i$ of the TBM matrix above need not vanish.
We call the phases $\phi_i$ in this basis as the ``mismatch phases'' 
to stay clear of the connotation of the word ``unphysical''.

TBM does by itself neither predict the Majorana or the unphysical phases 
nor the magnitudes or ordering of the neutrino masses.  
However, the magnitudes and ordering of the masses  
as well as the values of  
the Majorana phases are crucial for the size of renormalization group  
corrections \cite{antusch-majorana,more-recent,haba,DGR0} 
and the Planck scale corrections to this mixing scheme.  
Moreover, as we shall see in the course of the paper, in the analysis of  
Planck scale effects on tri-bimaximal mixing  
even the mismatch phases turn out to be crucial to modify any mixing  
angle.  
This is a feature specific to TBM, having to do with  
the structure of the corresponding neutrino mass matrix.  
In general, for a $\mu$--$\tau$ symmetric mass matrix \cite{mutau},  
corrections to $\theta_{13} = 0$ and $\theta_{23} = \pi/4$ require  
mismatch phases to take non-trivial values.  
 
Several interesting models giving rise to Eq.~(\ref{eq:Utri})  
have been proposed in the literature \cite{models}.   
The predictions of TBM, $\sin \theta_{13} = 0 = \cos 2 \theta_{23}$ and  
$\sin^2 \theta_{12} = 1/3$, may be compared with the current  
$3 \sigma$ ranges of the parameters \cite{thomas}:  
\bea \label{eq:data} 
0.24 \le \sin^2 \theta_{12} \le 0.40 ~,\\[0.2cm] 
-0.36 \le \cos 2 \theta_{23} \le 0.32~,\\[0.2cm] 
|\sin \theta_{13}| \equiv |U_{e3}| \le 0.20~, 
\eea 
with best-fit values of 0.31 for $\sin^2 \theta_{12}$ and zero  
for $\sin \theta_{13}$ and $\cos 2 \theta_{23}$.  
As far as the neutrino masses are concerned, 
oscillations experiments are sensitive only to the  
mass-squared differences, which are measured to be  
\bea 
7.1 \cdot 10^{-5}~{\rm eV}^2 \le \dms \equiv m_2^2 - m_1^2 \le  
8.9 \cdot 10^{-5}~{\rm eV}^2~,\\[0.2cm] 
1.9 \cdot 10^{-3}~{\rm eV}^2 \le \dma \equiv |m_3^2 - m_2^2| \le  
3.2 \cdot 10^{-3}~{\rm eV}^2~, 
\eea 
with the best-fit values of $\dms = 7.9 \cdot 10^{-5}~{\rm eV}^2$ and  
$\dma = 2.5 \cdot 10^{-3}~{\rm eV}^2$.  
Cosmological observations give an upper limit on the  
neutrino mass of $\ls 0.5$ eV \cite{steen}, which is stronger than the  
limits obtained by direct searches, $\le 2.3$ eV \cite{mainz}.  
It is still unknown whether neutrinos enjoy a normal hierarchy   
(NH: $m_3^2 \simeq \dma \gg m_2^2 \simeq \dms \gg m_1^2$), an  
inverted hierarchy  
(IH: $m_2^2 \simeq m_1^2 \simeq \dma \gg m_3^2$)  
or are quasi-degenerate  
(QD: $m_3^2 \simeq m_2^2 \simeq m_1^2 \equiv m_0^2 \gg \dma$).  
 
Many future planned/proposed experiments are geared towards  
improving the precision of the mixing angles and mass squared differences.  
Since this paper deals with the deviations of the  
mixing angles from the presently favored TBM scenario, we list  
some of the future experimental proposals which can be particularly  
suitable for this purpose.  
The precision of $\sin^2\theta_{12}$ can be considerably improved 
by a dedicated reactor neutrino experiment situated 
at $\sim$ 60 km corresponding to the survival probability  
minimum of $\bar{\nu}_e$   
\cite{th12,sado,th12new}. For instance it was shown in  
\cite{th12new} that with a statistics of $\sim$ 60 Giga-Watt kiloton year 
and a systematic error of 2\%, $\sin^2\theta_{12}$ can be measured  
to within $\sim$ 5\% at 3$\sigma$. The parameter  
$\sin^2\theta_{23}$ can be determined with an accuracy of $\sim$   
10\% at 3$\sigma$ depending on the true value of $\sin^2\theta_{23}$ 
in the T2K and No$\nu$a experiments \cite{tenyears,nova-proposal}.  
In what regards $\theta_{13}$,  
the Double Chooz experiment \cite{newexp} (see also \cite{tenyears}) will  
improve the $3\sigma$ limit on  
$|U_{e3}|$ from its current value  
0.04 to 0.01~(0.006) after 2~(6) years of data taking.

The outline of the paper is as follows:  
we start by discussing the RG running of TBM analytically in  
Section \ref{sec:RG}. 
In Section \ref{sec:Planck} we then turn to Planck  
scale effects and point out in particular the importance of the  
mismatch phases.  
Both types of corrections are discussed as functions of  
the neutrino mass and the type of ordering.  
A numerical analysis and comparison of  
both effects is performed in Section \ref{sec:num}.  
Section \ref{sec:concl} summarizes our findings.

\section{\label{sec:RG}Renormalization Effects on Tri-bimaximal Mixing}  
 
In this Section we will study the effect of renormalization group  
running of the mixing angles when at high scale they 
correspond to the tri-bimaximal mixing.  
 
RG effects on tri-bimaximal mixing have  
been studied before in Refs.~\cite{PR,xingRG,rabi,valle}, but an analysis  
involving all possible mass values and schemes is still lacking.  
We focus here mostly on the {\it maximal} RG effects as a function of the  
neutrino mass and will not conduct a detailed analysis of the  
influence of the Majorana phases on the running. This will be  
performed in a separate work \cite{DGR}. 
 
By inserting the neutrino mixing matrix from Eq.~(\ref{eq:Utri})  
in the definition of  
$m_\nu = U^* \, m_\nu^{\rm diag} \, U^\dagger$, one finds  
\be \label{eq:mnutri1} 
m_\nu =  
\left( 
\bad  
A \, e^{2 i \phi_1} & B \, e^{i (\phi_1 + \phi_2)}  
& B \, e^{i (\phi_1 + \phi_3)} \\[0.2cm] 
\cdot & \frac{1}{2} (A + B + D) \, e^{2 i \phi_2}  
& \frac{1}{2} (A + B - D) \, e^{i (\phi_2 + \phi_3)} \\[0.2cm] 
\cdot & \cdot & \frac{1}{2} (A + B + D)\, e^{2 i \phi_3}  
\ea  
\right)~.  
\ee 
This is the most general mass matrix generating tri-bimaximal mixing.  
The parameters $A, B$ and $D$ are given by  
\be \label{eq:ABD} 
A = \frac{1}{3} (2 \, m_1 \, e^{i \alpha_1} + m_2 \, e^{i \alpha_2})~,~~ 
B = \frac{1}{3} (m_2 \, e^{i \alpha_2} - m_1 \, e^{i \alpha_1})~,~~ 
D = m_3 \, e^{i \alpha_3}  \; .  
\ee 
The neutrino mass matrix is subject to RG evolution.  
The running of this matrix may be described via \cite{chankowski}  
\be \label{eq:mnuRG}  
m_\nu \rightarrow I_K \, I_\kappa \,  m_\nu \, I_\kappa \; ,  
\ee 
where $I_K$ is a flavor independent factor arising from  
the gauge interactions and fermion-antifermion loops.  
This factor does not influence the mixing angles at all.  
The diagonal matrix $I_\kappa$ is given by  
\be \label{eq:Ialpha} 
I_\kappa = {\rm diag}(e^{- \Delta_e},e^{- \Delta_\mu},e^{- \Delta_\tau})  
\simeq  
{\rm diag}(1,1,1 - \Delta_\tau) \; , 
\ee 
where 
\be 
\Delta_\tau \equiv \frac{m_\tau^2}{8 \pi^2 \, v^2}\, (1 + \tan^2 \beta)  
\, \ln \frac \Lambda\lambda~. 
\ee 
Here we have neglected the electron and muon mass  
(so that $\Delta_e=\Delta_\mu=0$)  
and used $e^{- \Delta_\tau} \simeq 1 - \Delta_\tau$,  
since $\Delta_\tau \ll 1$.  
For instance, for $\tan \beta = 20$ and $\Lambda/\lambda = 10^9$, one has  
$\Delta_\tau \simeq 0.0054$. The results for the SM can  
be obtained by replacing  
the factor $(1 + \tan^2 \beta)$ with $(-3/2)$.  
 
Note that the RG effects correspond to multiplying every entry of the  
neutrino mass matrix with a real number.  
Consequently, the overall phase of  
the entries is not affected by the corrections. The corresponding  
``unphysical phases'' $\phi_1$, $\phi_2$ and $\phi_3$ can therefore  
be rephased away by a redefinition of the charged lepton fields.  
Hence, for the analysis of RG  
effects it suffices to consider the mass matrix  
\be \label{eq:mnutri} 
m_\nu =  
\left( 
\bad  
A & B & B \\[0.2cm] 
\cdot & \frac{1}{2} (A + B + D) & \frac{1}{2} (A + B - D)\\[0.2cm] 
\cdot & \cdot & \frac{1}{2} (A + B + D) 
\ea  
\right)~.  
\ee 
Obviously, if the correction to the neutrino mass matrix is additive and  
not multiplicative, then the unphysical phases will play a role.  
We will show in the  
analysis of Planck scale effects to be presented in the next Section  
that this indeed is the case.

Returning to the RG effects, the mass matrix in (\ref{eq:mnutri}) is 
generated by some mechanism at a high scale $\Lambda$, which 
may be taken to be the typical seesaw scale  
$\Lambda \sim 10^{12}$ GeV. 
If the mechanism involves right-handed neutrinos, some of them are 
expected to have masses above $\Lambda$, and we have to assume 
that the threshold effects \cite{thresh} do not spoil the 
TBM relations till $\Lambda$ (note that additional unknown  
parameters, namely the entries of the Dirac neutrino mass matrix would  
enter the analysis). This will be a valid assumption, 
for example, when the heavy right-handed neutrinos are exactly degenerate. 
Anyway, the predictions of this mechanism will be modified by the RG  
evolution to the low energy  
scale $\lambda$ at which measurements take place.  
We take $\lambda = 10^3$ GeV, the typical scale of supersymmetry 
breaking. 
Note that the dependence on the actual values of the 
scales $\Lambda$ and $\lambda$ is only logarithmic, so our  
results are rather insensitive to the exact choice of scales. 
If the threshold effects indeed are sizeable, our results  
can be considered to be only conservative estimates.

In order to study the effect of the RG corrections on the mixing angles,  
we employ the strategy presented in  
\cite{DGR0} to diagonalize the RG-corrected mass matrix and  
obtain simple expressions for the evolved mixing angles  
\be \label{eq:tij_kij} 
\theta_{ij} \simeq \theta_{ij}^0 + k_{ij} \, \Delta_{\tau} + 
{\cal O}(\Delta_\tau^2) \; ,  
\ee 
where $\theta_{12}^0 = \arcsin\sqrt{1/3}$, $\theta_{13}^0 = 0$ and  
$\theta_{23}^0 = -\pi/4$. We only quote the result \cite{DGR}:  
\bea \label{eq:kij} 
\D k_{12} = \frac{1}{3\sqrt{2}} 
\, \frac{\left| m_1 + m_2 \, e^{i \alpha_2}\right|^2}{\dms} ~,\\[0.2cm] 
\D k_{23} = - 
\left(  
\frac{1}{3} \, \frac{\left| m_2 + m_3 \, e^{i(\alpha_3 - \alpha_2)} 
\right|^2}{m_3^2 - m_2^2}   
+ \frac{1}{6} \, \frac{\left| m_1 + m_3 \, e^{i\alpha_3}\right|^2} 
{m_3^2 - m_1^2}   
\right) ~,\\[0.2cm] \D  
k_{13} = -\frac{1}{3\sqrt{2}} 
\left(  
\frac{\left| m_2 + m_3 \, e^{i(\alpha_3 - \alpha_2)} 
\right|^2}{m_3^2 - m_2^2}   
- \frac{\left| m_1 + m_3 \, e^{i\alpha_3}\right|^2} 
{m_3^2 - m_1^2}   
\right) ~. 
\eea 
Using the additional rephasing freedom we have set $\alpha_1 = 0$.   
Note that as long as the RG evolutions of the angles are much less than  
${\cal O}(1)$,   
the difference in all the $m_i/m_j$ ratios at the low and high scales 
is only ${\cal O}(\Delta_\tau)$ \cite{DGR0}.  
Therefore for ${\cal O}(\Delta_\tau)$ estimates, 
one may use the $m_i$ values at the low scale. 
In the numerical calculations we have performed, the RG evolution of the  
$m_i$ \cite{antusch-majorana} are also taken into account.

It may be noted that for $k_{12}$ there  
is no dependence on the third mass eigenstate or its  
Majorana phase $\alpha_3$.  
As expected, $k_{13}$ and $k_{23}$ are governed by the  
inverse of $\dma$ while $k_{12}$ depends on  
$1/\dms$, which renders it typically larger. As $k_{12}$ is always positive,  
the solar neutrino mixing angle always increases in the MSSM and decreases  
in the SM. In contrast, $k_{23}$ is negative (positive) for  
a normal (inverted) mass ordering. 
Therefore $|\theta_{23}|$ increases in the MSSM and decreases  
in the SM if the neutrino mass ordering is normal.  
If the ordering is inverted,  
$|\theta_{23}|$ decreases in the MSSM and increases in the SM.  
From $\theta_{ij} = \theta_{ij}^0 + k_{ij} \, \Delta_{\tau}$ it  
follows for $|k_{ij} \, \Delta_\tau | \ll 1$ that  
\be \label{eq:obs} 
\sin \theta_{13} \simeq k_{13} \, \Delta_\tau~ , ~ 
\cos 2 \theta_{23} \simeq 2 \, k_{23} \, \Delta_\tau~,~ 
\sin^2 \theta_{12} -\frac{1}{3} \simeq  
\frac{2 \sqrt{2}}{3} \, k_{12} \, \Delta_\tau ~. 
\ee 
In the next Subsections we will discuss from these expressions  
the features of the running for the three main types of 
neutrino mass spectra:  
normal (NH) and  inverted (IH) hierarchy, and quasi-degeneracy 
(QD).  
In order to give numerical estimates of the extent of RG effects,  
we use the best-fit values of  
$\dms = 7.9 \cdot 10^{-5}$ eV$^2$, $\dma = 2.5 \cdot 10^{-3}$ eV$^2$,  
and choose the phases such that the RG effect is maximized.  
Then we further choose $\tan \beta = 20$ for illustration.   
The results for the SM can be obtained by replacing  
the factor $(1 + \tan^2 \beta)$ with $(-3/2)$.  
The outcome of a full numerical analysis is plotted in Figs.~\ref{theta13}, 
\ref{theta23} and \ref{theta12}: as we shall see, they are nicely  
reproduced by the analytical estimates to be presented below.

\subsection{Normal Hierarchy}

We start the estimates with the normal hierarchy.  
Defining the notation 
\be 
r \equiv \sqrt{\dms/\dma} \simeq 0.18 ~, 
\ee 
we have $m_1 \simeq 0$, $m_2 \simeq r \sqrt{\dma}$  
and $m_3 \simeq \sqrt{(1 + r^2) \, \dma}$.  
From Eqs.~(\ref{eq:kij}, \ref{eq:obs}) we get 
\be \label{eq:t13NH} 
|\sin\theta_{13}|_{\rm NH} \simeq 
\frac{\sqrt{2}}{3} \, \Delta_\tau \,  
\left| r \, \cos (\alpha_2 - \alpha_3) + r^2 \right| 
 \ls 1.4 \cdot 10^{-6}~(1 + \tan^2 \beta)  
\rightarrow 5.4 \cdot 10^{-4} \; , 
\ee 
where the number indicated by the arrow is the value at  
$\tan\beta=20$. 
Atmospheric neutrino mixing is now non-maximal, the mixing angle 
being given by   
\begin{eqnarray}\label{eq:t23NH} 
(\cos 2 \theta_{23})_{\rm NH} & \simeq & - \Delta_\tau \,  
\left(1 + \frac 43 \, r \, \cos (\alpha_2 - \alpha_3)  
\right) \nonumber \\ 
\Rightarrow |\cos 2 \theta_{23}|_{\rm NH} &  
\ls & 1.7 \cdot 10^{-5} ~(1 + \tan^2 \beta)  
\rightarrow 6.8 \cdot 10^{-3} ~. 
\end{eqnarray} 
Note that $(\cos 2 \theta_{23})_{\rm NH}<0$. 
The solar neutrino mixing angles increases from $\sin^2 \theta_{12} = \frac 13$  
by   
\be\label{eq:t12NH} 
\left(\sin^2 \theta_{12} - \frac 13\right)_{\rm NH}  \simeq  
\frac{2}{9} \, \Delta_\tau  \simeq 3.0 \cdot 10^{-6} ~(1 + \tan^2 \beta)  
\rightarrow 1.2 \cdot 10^{-3}~.  
\ee 
Thus, we find that for NH,  
the deviations of the angles from tri-bimaximal values  
due to RG corrections are extremely small and 
virtually impossible to probe. 
Note that all the three deviations as plotted in the upper panels  
of Figs.~\ref{theta13}, \ref{theta23} and \ref{theta12} are  
independent  
of the value of $m_1$ as long as $m_1 \ll \sqrt{\dms} \simeq 9 \cdot  
10^{-3}$ eV.

\subsection{Inverted Hierarchy} 
 
Turning to the inverted hierarchy,  
using $m_2 = \sqrt{m_3^2 + \dma}$ and $m_1 = \sqrt{m_3^2 + 
(1-r^2) \, \dma}$,  
it follows from Eqs.~(\ref{eq:kij}, \ref{eq:obs}) that  
\begin{eqnarray} \label{eq:t13IH} 
|\sin \theta_{13}|_{\rm IH} & \simeq &  
\frac{m_3}{\sqrt{\dma}} \, \frac{\sqrt{2}}{3} \, \Delta_\tau \,  
\left| \cos(\alpha_2 - \alpha_3) - \cos \alpha_3 \right| \nonumber \\ 
 & \ls &  2.6 \cdot 10^{-7} \, \left(\frac{m_3}{10^{-3}~\rm eV} 
\right) \, (1+\tan^2 \beta)  
\to 1.0 \cdot 10^{-4} \left(\frac{m_3}{10^{-3}~\rm eV}\right) \; . 
\end{eqnarray} 
Thus in the inverted hierarchy, the value of $\sin \theta_{13}$ generated 
through RG evolution 
is proportional to $m_3$, as can be seen in the lower panel of  
Fig.~\ref{theta13}.  
The atmospheric neutrino mixing angle is given by  
\be\label{eq:t23IH} 
(\cos 2 \theta_{23})_{\rm IH} \simeq   \Delta_\tau  
\simeq 1.4 \cdot 10^{-5} ~(1 + \tan^2 \beta)  
\rightarrow 5.4 \cdot 10^{-3} ~, 
\ee 
which is independent of $m_3$ as long as $m_3 \ll \sqrt{\dma}  
\simeq 0.05$ eV. 
The dependence on the Majorana phases is introduced only at  
order $m_3/\sqrt{\dma}$.  
As alluded to before, in case of a normal ordering we have  
$(\cos 2\theta_{23})_{\rm NH} <0$, while for an inverted ordering  
$(\cos 2\theta_{23})_{\rm IH} >0$, i.e., RG effects increase 
$|\theta_{23}|$ from its original TBM value of $\pi/4$ 
for a normal hierarchy and decrease it for an  
inverted hierarchy. In the SM the effects have the opposite sign.  
Finally, the solar neutrino mixing angle increases for the MSSM and reads  
\be \label{eq:t12IH}   
\left(\sin^2 \theta_{12} - \frac 13\right)_{\rm IH}  \simeq  
\frac{4 \Delta_\tau}{9 \, r^2} \, (1 + \cos \alpha_2)  
\ls 3.9 \cdot 10^{-4} ~(1 + \tan^2 \beta)  
\rightarrow 0.15  ~. 
\ee 
Even though Eq.~(\ref{eq:tij_kij}) is strictly speaking no longer valid  
for such large values of $|\theta_{ij}-\theta_{ij}^0|$,  
the above result indicates that the running of the solar neutrino  
mixing angle can be dramatic, namely up to $10^\circ$.  
This will be confirmed in Sec.~\ref{sec:num}, where  
we will present figures quantifying the maximal RG effect, which are  
obtained by numerically solving the RG equations from  
Ref.~\cite{antusch-majorana}.  
 
The leading term in Eq.~(\ref{eq:t12IH}) is suppressed when  
$\alpha_2 \simeq \pi$.  
The measured value of $|\theta_{12} - \theta_{12}^0| < 4^\circ$  
(or $\sin^2 \theta_{12} - \frac 13 \le 0.07$)  
at 3$\sigma$ suggests that the observed value of $\theta_{12}$ can be  
used to constrain the values of the absolute neutrino masses, the  
Majorana phase $\alpha_2$ and $\tan\beta$ in IH \cite{PR,DGR}.  
In contrast to this,  
$\sin\theta_{13}$ and $\cos 2 \theta_{23}$ are too small to be observable.

\subsection{Quasi-degeneracy} 
 
Finally, we shall consider running for quasi-degenerate neutrinos.  
The RG running for QD is always the largest: 
indeed, the deviation of angles from their  
tri-bimaximal values grows quadratically with the common mass scale $m_0$.  
Using $m_0^2 \gg \dma$ for the approximations and  
choosing $m_0 = 0.2$ eV for 
illustration, it follows that  
\be \label{eq:t13QD} 
|\sin\theta_{13}|_{\rm QD} \simeq \frac{\sqrt{2}}{3} \, \Delta_\tau \,  
\frac{m_0^2}{\dma} \,  
\left| \cos (\alpha_2 - \alpha_3) - \cos \alpha_3 \right| 
\ls 2.0 \cdot 10^{-4}~(1 + \tan^2 \beta)  
\rightarrow 0.08 ~. 
\ee 
Thus, testable values of $\theta_{13}$ up to $5^\circ$ may be generated.  
Turning to atmospheric mixing, one finds 
\begin{eqnarray} \label{eq:t23QD}  
(\cos 2 \theta_{23})_{\rm QD} & \simeq & \mp \frac 23 \, \Delta_\tau \,  
\frac{m_0^2}{\dma} \,  
\left(3 + 2 \,  \cos (\alpha_2 - \alpha_3) + \cos \alpha_3 \right)  
\nonumber \\  
\Rightarrow |\cos 2 \theta_{23}|_{\rm QD} 
& \ls & 8.0 \cdot 10^{-4} ~(1 + \tan^2 \beta)  
\rightarrow  0.32~, 
\end{eqnarray} 
where the $-$ sign is for normal ordering and the $+$ sign for  
inverted ordering. The running is maximal  
when all the Majorana phases vanish. 
The value of $\theta_{23}$ can deviate from its maximal  
value by up to $10^\circ$.  
This deviation is currently restricted by experiments to 
$|\theta_{23} - \theta_{23}^0| < 10^\circ$ at 3$\sigma$. 
Therefore, more accurate future measurements of $\theta_{23}$  
can be used to put bounds on $m_0, \alpha_2$ and $\alpha_3$ \cite{DGR}. 
The value of $\alpha_2$ may be restricted to $\alpha_2 \simeq 
\pi$ from the $\theta_{12}$ measurements as we shall see next:  
the solar mixing angle can deviate strongly from its tri-bimaximal 
value:  
\be \D \label{eq_t12QD} 
\left(\sin^2 \theta_{12} - \frac 13 \right)_{\rm QD} \simeq  
\frac 49 \, \Delta_\tau \, (1 + \cos \alpha_2) \,  
\frac{m_0^2}{\dms}  
\simeq 3.0 \cdot 10^{-3} ~(1 + \tan^2 \beta) \, (1 + \cos \alpha_2)  
~. 
\ee 
Since the maximum value of the quantity  
$\left(\sin^2 \theta_{12} - \frac 13 \right)$ can be 0.67 one can obtain a 
condition\footnote{The same comments as given after Eq.~(\ref{eq:t12IH})  
apply here.}  
for the validity of the analytic expressions as $(m_0/{\rm eV})  
\, \tan\beta \lsim 4$. 
At large $\tan\beta$, the quantity $\theta_{12}$ can be too large to be  
accommodated 
by the data, unless $\alpha_2 \simeq \pi$.  
This simplifies the predictions for the other two angles: 
\begin{eqnarray}  
\label{eq:QDcons} 
|\sin\theta_{13}|_{\rm QD}^{\alpha_2 = \pi}  
& \simeq & \frac{2\sqrt{2}}{3} \, \Delta_\tau \,  
\frac{m_0^2}{\dma} \, |\cos \alpha_3| \ls  
2.0 \cdot 10^{-4} \,(1 + \tan^2 \beta) \rightarrow 0.08 ~,\\ 
|\cos 2 \theta_{23}|_{\rm QD}^{\alpha_2 = \pi}  
& \simeq & 2 \, \Delta_\tau \, \frac{m_0^2}{\dma} \,  
\left| 1 - \frac 13 \, \cos \alpha_3 \right| \ls 
 5.4 \cdot 10^{-4} \,(1 + \tan^2 \beta) \rightarrow 0.22~, 
\end{eqnarray} 
with the sign of $\cos 2\theta_{23}$ negative (positive) for 
normal (inverted) mass ordering. 
Note that $\sin\theta_{13}$ can be zero if  
$\alpha_3 = \pi/2$, whereas $\cos 2 \theta_{23}$ is  always non-zero. 
  
Another phenomenological implication of $\alpha_2 \simeq \pi$ and  
quasi-degenerate neutrinos is that the  
effective mass $\meff = |\sum U_{ei}^2 \, m_i|$  
governing \onbb \cite{0vbb} takes its minimal possible value:  
\be 
\meff_{\rm QD}^{\alpha_2 = \pi} \simeq  m_0 \, \cos 2 \theta_{12}~ 
\ee 
in the $\theta_{13} \to 0$ limit. 
The same formula with $m_0$ replaced by $\sqrt{\dma}$ is valid for  
an inverted hierarchy and $\alpha_2 \simeq \pi$, which suppresses  
the running also in this case (see Eq.~(\ref{eq:t12IH})).

\section{\label{sec:Planck}Planck Scale Effects on Tri-bimaximal Mixing} 
 
In this Section we will discuss the implications of Planck scale  
physics on tri-bimaximal mixing.  
The existence of the  
Planck scale implies the presence of higher  
dimensional non-renormalizable interactions,  
among which the following dimension five  
operator is of interest for neutrino physics:  
\be 
{\cal L}_{\rm Gr} = \frac{\lambda_{\alpha \beta}}{M_{\rm Pl}} \,  
\left(L_\alpha \, \Phi \right) \, \left(L_\beta \, \Phi \right)  
+ h.c.  
\ee 
The coupling matrix $\lambda_{\alpha \beta}$ will be assumed  
to be flavor democratic, since gravity does not distinguish  
between flavors.  
After electroweak symmetry breaking  
the above operator leads to a contribution to the  
low energy neutrino mass matrix $m_\nu \rightarrow  
m_\nu + \delta m_\nu$ of the form  
\be \label{eq:mnudelta} 
\delta m_\nu = \mu \,  
\left( 
\bad  
1 & 1 & 1 \\[0.2cm] 
1 & 1 & 1 \\[0.2cm] 
1 & 1 & 1  
\ea 
\right) \equiv \mu \, \Delta \mbox{ with }  
\mu \simeq  \frac{\langle \Phi \rangle^2}{M_{\rm Pl}} \simeq 2.5 \cdot 10^{-6}~\rm eV~. 
\ee 
The implications of such a correction to $m_\nu$ have been  
noted and analyzed for instance in Refs.~\cite{planck,ue3,new}.  
With $\mu \simeq 2.5 \cdot 10^{-6}~ {\rm eV} \ll \sqrt{\dms}$, 
it would appear that  
only negligible corrections to the mixing phenomena can be expected. 
However, as will be seen later in this section, the Planck scale 
effects on $\theta_{12}$ are governed by $\mu \, m_0 /\dms$, 
which can be substantial for quasi-degenerate neutrinos. 
Moreover we stress that the presence of such a term is expected  
on general  
grounds and its implications are therefore model independent.  
In order to consider all possible corrections to a given mixing 
scheme, the perturbation (\ref{eq:mnudelta}) has to be included. 

Note that the Planck scale contribution to the neutrino mass matrix, 
$\delta m_\nu$, has all its elements real only with a specific choice
of the phases of the flavor eigenstates. In this basis, the 
so-called ``unphysical'' phases of the TBM matrix 
(see Eq.~(\ref{eq:Utri})) need not vanish.
These ``mismatch phases'' -- the phases $\phi_i$ in this 
particular basis --
turn out to be crucial in the context of the tri-bimaximal mixing,
since without them there would be no effect on the mixing angles at all.  
To show this, consider the neutrino mass matrix giving rise to TBM in the  
absence of the $\phi_{1,2,3}$. It is given in Eq.~(\ref{eq:mnutri})  
and can be written as:  
%
% 
%We have noted before (after Eq.~(\ref{eq:mnutri}))  
%that the presence of an additive correction to  
%the neutrino mass matrix renders the ``unphysical'' phases $\phi_{1,2,3}$  
%(see Eq.~(\ref{eq:Utri})) to be no longer ``unphysical''.   
%In the context of tri-bimaximal mixing these phases turn out to be crucial,  
%since without them there would be no effect on the mixing angles at all.  
%To show this, consider the neutrino mass matrix giving rise to TBM in the  
%absence of the $\phi_{1,2,3}$. It is given in Eq.~(\ref{eq:mnutri})  
%and can be written as:  
\be \label{eq:interms} 
m_\nu = \frac{m_1}{6} \, e^{i \alpha_1} \,  
\left(  
\bad  
4 & -2 & -2 \\ 
\cdot & 1 & 1 \\ 
\cdot & \cdot & 1  
\ea  
\right)  
+ \frac{m_2 }{3} \, e^{i \alpha_2} \,  
\left(  
\bad  
1 & 1 & 1 \\ 
\cdot & 1 & 1 \\ 
\cdot & \cdot & 1  
\ea  
\right) +  
\frac{m_3 }{2} \, e^{i \alpha_3} \,  
\left(  
\bad  
0 & 0 & 0 \\ 
\cdot & 1 & -1 \\ 
\cdot & \cdot & 1  
\ea  
\right)~. 
\ee 
The neutrino with mass $m_2$ has a flavor democratic  
contribution to the total mass matrix. This is exactly the flavor-blind  
form that the Planck scale contribution $\delta m_\nu$ in  
Eq.~(\ref{eq:mnudelta}) possesses.  
Hence, adding $\delta m_\nu$ to Eq.~(\ref{eq:interms}) is  
equivalent to a simple redefinition of $m_2$ as 
\be 
m_2 \, e^{i \alpha_2} \longrightarrow  
m_2 \, e^{i \alpha_2} \, \left( 1 + \epsilon_\mu \, e^{-i \alpha_2}\right)~, 
\ee 
where $\epsilon_\mu = 3 \, \mu/m_2 \ll 1$.  
Hence, only the value of the second neutrino mass and its corresponding  
Majorana phase are modified, while the mixing angles and the other  
masses remain unchanged: with $\mu \simeq 2.5 \cdot 10^{-6}$  
eV and $m_2 \ge 8.4 \cdot 10^{-3}$ eV, it holds that  
$\epsilon_\mu \ls 8.9 \cdot 10^{-4}$ and the effect on $m_2$  
and on the mass squared differences is at most of the order of 
0.1\%.

However, in general the situation is different since  
the most general mass matrix giving rise to TBM,  
as given in Eq.~(\ref{eq:mnutri1}), contains  
the phases $\phi_i$.  
The new addition of a flavor democratic small  
perturbation to the neutrino mass matrix cannot be compensated anymore  
by a redefinition of $m_2 \, e^{i \alpha_2}$. This is  
because it is not possible to write $m_\nu$ in terms of the  
individual masses as in Eq.~(\ref{eq:interms}),  
i.e., the elements of the matrix multiplying $m_2 e^{i\alpha_2}$  
are complex numbers. 
The corrections to TBM from the Planck scale effects 
are then nontrivial\footnote{ 
It has been noted  
\cite{noted} that the tri-bimaximal mixing scenario and  
Quark-Lepton Complementarity scenarios \cite{QLC}, which link the  
CKM and PMNS matrices, generate basically  
the same $\sin^2 \theta_{12}$. We remark here that the QLC scenarios  
will be affected by Planck scale effects even if  
the phases $\phi_i$ vanish.}, 
and henceforth we shall consider this general scenario. 
Note that though the vanishing of the mismatch phases is a special case,  
it may be relevant while postulating symmetries at the high scale 
that govern the structure of the mass terms.\\ 
 
We would like to make a few remarks at this stage:  
\begin{itemize} 
\item 
The so-called ``unphysical'' phases, that are usually absorbed 
in the phases of charged leptons while constructing the leptonic
mixing matrix, indeed are not well-defined
outside the context of a theory of neutrino masses.
However, a complete theory of neutrino masses has to predict 
the magnitudes and phases of all the terms in the neutrino 
mass matrix, and hence in the mixing matrix $U$ from 
Eq.~(\ref{eq:Utri}),
once a choice for the phases of neutrino flavor eigenstates
has been made.
Such a choice has been made while writing the real democratic
matrix $\delta m_\nu$ in Eq.~(\ref{eq:mnudelta}). 
Hence it should not be surprising that some of the 
predictions of the theory -- like the values of the
mixing angles -- do indeed depend on the mismatch phases;

%in the absence of the Planck scale contribution $\delta m_\nu$ 
%to the neutrino mass matrix, the TBM matrix has the standard form  
%of Eq.~(\ref{eq:Utri}) with the ``unphysical'' phases $\phi_i=0$ 
%only with the charged lepton phases chosen specifically for  
%this purpose. 
%When the Planck scale contributions are added, the charged 
%lepton phases that need to be chosen are different in general. 
%This mismatch of phases is physical, and therefore the 
%``unphysical'' phases play a role in determining the extent 
%of the Planck scale effects \cite{ue3};  

\item 
The argument presented here regarding the vanishing of  
Planck effects on all three mixing angles and the masses $m_{1,3}$  
with vanishing mismatch phases is specific  
to the TBM scenario, where the mass matrix includes a flavor blind term.  
However, similar results based on symmetries can be obtained in  
more general cases. 
For example, TBM mixing is a special case of $\mu$--$\tau$ 
symmetry of the mass matrix, which implies the form  
(including the mismatch phases) \cite{mutau}:  
\be \label{eq:mutau} 
m_\nu^{\mu-\tau} =  
\left(  
\bad 
A \, e^{2 i \phi_1} & B \, e^{i (\phi_1 + \phi_2)}  
& B \, e^{i (\phi_1 + \phi_3)} \\ 
\cdot & D \, e^{2 i \phi_2} & E \, e^{i (\phi_2 + \phi_3)} \\  
\cdot & \cdot & D \, e^{2 i \phi_3}  
\ea 
\right)~, 
\ee  
where $A, B, D, E$ define $\theta_{12}$, the neutrinos masses and  
Majorana phases. Due to the equality of the $e\mu$ and $e\tau$ elements,  
as well as of the $\mu\mu$ and $\tau\tau$ elements, maximal  
atmospheric mixing and vanishing $U_{e3}$ results.   
Note now that the Planck scale contribution $\delta m_\nu$ from  
Eq.~(\ref{eq:mnudelta}) is also $\mu$--$\tau$ symmetric. Hence,  
adding $\delta m_\nu$  
to the $\mu$--$\tau$ symmetric matrix in case of vanishing  
$\phi_{1,2,3}$ will keep the total mass matrix $\mu$--$\tau$ symmetric  
and the values $U_{e3} = \cos 2 \theta_{23} = 0$ will not be changed.  
In order for the Planck effects to change the values of these 
angles the mismatch phases are required not to vanish; 
%if the matrix $\delta m_\nu$ is democratic and real.
 
\item  
the net order of magnitude of Planck effects should 
stay the same even if we take the elements of the democratic 
matrix $\delta m_\nu$ from  
Eq.~(\ref{eq:mnudelta}) to be ${\cal O}(1)$, and not necessarily exactly  
equal. 
 
\end{itemize}

Now we will consider the general case of all possible phases in the neutrino  
mass matrix and study the resulting effect of the Planck scale contribution  
for the mixing angles in case of TBM.  
Towards this end, we follow the formalism of \cite{ue3} to  
calculate the deviations  
of the mixing angles from their TBM values. 
We use the shorthand notation 
\be 
M_{ij} = \mu \, (U^T \, \Delta \, U)_{ij}~, 
\ee 
such that $M_{ij}$ incorporates all the $\phi_i$ dependence 
and a part of the $\alpha_i$ dependence.  
In general, the elements of $M_{ij}$ are ${\cal O}(\mu)$.  
To a good approximation, the mass squared difference  
$m_3^2 - m_1^2$ does not change  
due to the Planck scale effects. 
In this limit one finds that \cite{ue3}  
\begin{eqnarray}  
\label{eq:delue3} 
\delta U_{e3}  & \simeq&   \sum\limits_{i=1,2}  
\, U_{ei} \,  
\left(\frac{ \Re (M_{i3})}{m_3 \, e^{i \alpha_3} - m_i \, e^{i \alpha_i}}  
- i \, \frac{\Im (M_{i3})}{m_3 \, e^{i \alpha_3} + m_i \, e^{i \alpha_i}} 
\right) ~, \\ 
\label{eq:delt23} 
\delta U_{\mu 3} & \simeq & \sum\limits_{i=1,2}  
\, U_{\mu i} \,  
\left(\frac{ \Re (M_{i3})}{m_3 \, e^{i \alpha_3} - m_i \, e^{i \alpha_i}}  
- i \, \frac{\Im (M_{i3})}{m_3 \, e^{i \alpha_3} + m_i \, e^{i \alpha_i}} 
\right) ~. 
\end{eqnarray} 
These quantities can be related to the Planck-corrected mixing angles after  
electroweak symmetry breaking through 
\begin{eqnarray} 
|\sin\theta_{13}| &  
=  
&  
|\delta U_{e3}| ~,\\[0.2cm] 
\cos 2\theta_{23}  
&  
=  
&  
1 - 2 \, \sin^2 \theta_{23}  
\simeq 1 - 2 \, |U_{\mu 3} + \delta U_{\mu 3}|^2  
\simeq   
- 2 \sqrt{2} \, |\delta U_{\mu 3}| \, \cos \chi_{\mu 3}~, 
\end{eqnarray} 
where $\chi_{\mu 3}$ is the relative phase between 
$U_{\mu 3}$ and $\delta U_{\mu 3}$. We see that the deviation from maximal  
atmospheric neutrino mixing can go in either direction.

We can estimate the size of the Planck scale contribution to  
be at most of the order $\mu/(m_3 - m_1)$ for $|\delta U_{e3}|$ and  
$|\delta U_{\mu3}|$.  
For neutrinos with a normal hierarchy and $m_1 = 0$, this  
quantifies to $\mu/\sqrt{\dma} \simeq 6 \cdot 10^{-5}$, 
and is thus negligibly small. 
An inverted hierarchy with $m_3=0$ gives the same result.   
The largest effect can be expected for quasi-degenerate neutrinos,  
in which case the corrections are  
at most of order $\mu \, m_0/\dma \simeq 5 \cdot 10^{-4}$. 
We have  
inserted here for illustration a value of $m_0 = 0.4$ eV for the  
common mass scale.  
Note that the corrections here are proportional to $m_0$, as 
opposed to $m_0^2$ in the case of the RG running. 
 
The modifications of the mixing angles  
depend on the values of the Majorana phases (as for radiative  
corrections) but in particular also on the mismatch phases.  
The full expressions for  
Eqs.~(\ref{eq:delue3}, \ref{eq:delt23}) are rather  
lengthy and not very instructive.  
An example we give is for the normal hierarchy, in which case  
one can roughly estimate  
\be \label{eq:ex1} 
|\delta U_{e 3}| \simeq  
\sqrt{2} \, \left|\sin \left(\frac{\phi_2 - \phi_3}{2} \right) \right|  
\frac{\mu}{\sqrt{\dma}} ~\mbox{ and } ~ 
|\delta U_{\mu 3}| \simeq  
\frac{|\sin (\phi_2 - \phi_3)|}{\sqrt{2}} \frac{\mu}{\sqrt{\dma}}~. 
\ee 
The Majorana phases appear only at ${\cal O}(r \, \mu /\sqrt{\dma})$.  
These expressions show explicitly that  
for vanishing mismatch phases $\phi_i$ the  
corrections to the mixing angles vanish. 
  
As in the case of the radiative corrections, the corrections to 
the mixing angles $\theta_{ij}$ are approximately 
inversely proportional to the mass squared difference 
$\Delta m^2_{ij}$,  
and therefore one may expect  
a more sizable correction to $\theta_{12}$  
than to $\theta_{13}$ and $\theta_{23}$.  
The contribution from the Planck scale to $U_{e2}$ is \cite{ue3}  
\be \label{eq:delt12} 
\delta U_{e 2} \simeq  \frac{U_{e 1}}{\dms} \,  
\left( \Re (M_{12}) ~ |m_1 + m_2 \, e^{i\alpha_2}| 
- i \, \Im (M_{12}) ~ |m_1 - m_2 \, e^{i\alpha_2}| 
\right)~, 
\ee 
where there is only one term because for TBM $U_{e3}=0$ 
holds before the Planck scale effects are included.  
Note that in the above expression, the $m_i$ and  
$\alpha_i$ are defined before the Planck scale effects are  
switched on, whereas $\dms$ is taken to be after the 
Planck scale effects are switched on. 
If $\dms$ does not change much due to the Planck scale effects, 
then $|m_1 \pm m_2 \, e^{i\alpha_2}|/\dms \simeq 
1/|m_1 \mp m_2 e^{i\alpha_2}|$ 
and $\delta U_{e2}$ could be written in the same form as  
the expressions for $\delta U_{e3}$  
and $\delta U_{\mu 3}$. For the analytical estimates, we will assume  
that this is the case. The numerical analysis in the  
next Section will not make this assumption. However,  
the expressions obtained here are quite close to the full result.

Since $\sin\theta_{12} \simeq |U_{e2}+\delta U_{e2}|$, the quantity 
in Eq.~(\ref{eq:delt12}) can be related to the  
deviation of the solar mixing angle via 
\be 
\sin^2 \theta_{12} - 1/3 \simeq |U_{e2} +\delta U_{e2}|^2 - 1/3 
\simeq \frac{2}{\sqrt{3}} \, |\delta U_{e2}| \, \cos \chi_{e2}~, 
\label{eq:chie2} 
\ee 
where $\chi_{e2}$ is the relative phase between 
$U_{e2}$ and $\delta U_{e2}$.  
The deviation can thus have either sign 
as in the case of the atmospheric neutrino mixing angle,

If initially $m_1=0$ (normal hierarchy), one can estimate  
$|\delta U_{e2}| \ls \mu/\sqrt{\dms} \simeq 3 \cdot 10^{-4}$.  
An inverted hierarchy leads to a larger correction,  
$|\delta U_{e2}| \ls 2 \, \mu \, \sqrt{\dma}/\dms \simeq 4 \cdot 10^{-3}$.  
Hence, in contrast to the corrections to  
$\theta_{13}$ and $\theta_{23}$, the deviation of  
solar neutrino mixing from 1/3 is sensitive to whether neutrinos are  
normally or inversely ordered.  
For quasi-degenerate neutrinos it follows  
from Eq.~(\ref{eq:delt12}) that  
$|\delta U_{e2}| \simeq \mu \, m_0/ \dms \simeq 10^{-2}$,  
where we have again used $m_0 = 0.4$ eV.
With $m_0^2 \gg \dms$, we can write 
%To give an analytical example we set the Majorana phases to zero  
%and find   
\bea 
|\delta U_{e2}| \simeq \D \frac{2}{3\sqrt{3}} \,  
\frac{\mu \, m_0 }{\dms}  \\[0.2cm] 
\left| 2 \cos 2 \phi'_1 - \cos 2 \phi'_2 + \cos (\phi'_1 + \phi'_2) -  
\cos 2 \phi'_3 + \cos (\phi'_1 + \phi'_3) - 2 \, 
\cos (\phi'_2 + \phi'_3) \right|~, 
\eea 
where $\phi'_i \equiv \phi_i - \alpha_2/4$.
This quantity, as it should, vanishes for $\phi_{1,2,3}=0$. The  
function of the $\phi_{1,2,3}$ inside the $|...|$ above 
has a maximal value of $\simeq 6.36$  
and therefore $|\delta U_{e2}| \ls 2.5 \, \mu \, m_0/\dms \simeq 0.03$  
for $m_0 = 0.4$ eV.  
Thus, with the maximal deviation being  
$\sin^2 \theta_{12} - \frac 13 \simeq (2/\sqrt{3}) \, |\delta U_{e 2}|$  
it follows that the deviation of $\theta_{12}$ from its 
TBM value due to Planck scale effects can be up to 10 \% or  
$2^\circ$--$3^\circ$.   
Moreover, as shown in Eq.~(\ref{eq:chie2}), this deviation can be in either 
direction depending on the relative phase $\chi_{e2}$,  
which in turn depends on  
the Majorana phases $\alpha_i$ as well as on the mismatch
phases $\phi_i$.

If we now want to incorporate both the RG as well as the Planck scale 
effects on $\theta_{12}$, 
the predictions are thus uncertain by $2^\circ$--$3^\circ$ 
in the absence of any knowledge of the $\phi_i$. 
This would relax some of the constraints on $m_0, \alpha_2$ 
and $\tan\beta$ that have been obtained in the literature 
by requiring the angles at the low scale to be compatible 
with the experiments. To be more quantitative, the maximal deviation  
from the initial value of $\theta_{12} = \arcsin \sqrt{1/3}$  
for a normal ordering is below $0.1^\circ$  
for $m_1 < 0.02$ eV. For $m_1 = 0.1$ eV the change can  
be up to $0.6^\circ$ and then it increases linearly with the neutrino mass,  
for instance $3^\circ$ for $m_1 = 0.5$ eV.  
If neutrinos are inversely ordered, the change is  
up to $0.3^\circ$ even for a vanishing $m_3$, and starts to increase  
linearly with the neutrino mass for $m_3 \gs 0.01$ eV.  
If radiative corrections  
are used to set constraints on the parameters, then one should  
take this uncertainty into account, which would weaken  
the corresponding limits.

One might wonder at this point  
whether one can generate successful phenomenology starting with  
bimaximal neutrino mixing \cite{bimax}, i.e.,  
with Planck scale contributions perturbing an  
initial\footnote{See also \cite{new} for a discussion on Planck  
scale effects on bimaximal mixing.} $\sin^2 \theta_{12} = \frac 12$.  
We checked numerically that in order to obey the current $3\sigma$ limit of  
$\sin^2 \theta_{12} \leq 0.4$, neutrinos should be  
heavier than 1.4 eV, i.e., in conflict with the already very tight neutrino  
mass limits from cosmology.  
Since the RG effects tend to increase $\theta_{12}$, even their 
addition cannot salvage the scenario.

\section{\label{sec:num}Numerical Results and Discussion}  
 
\begin{figure}[hbt] 
\begin{center} 
\epsfig{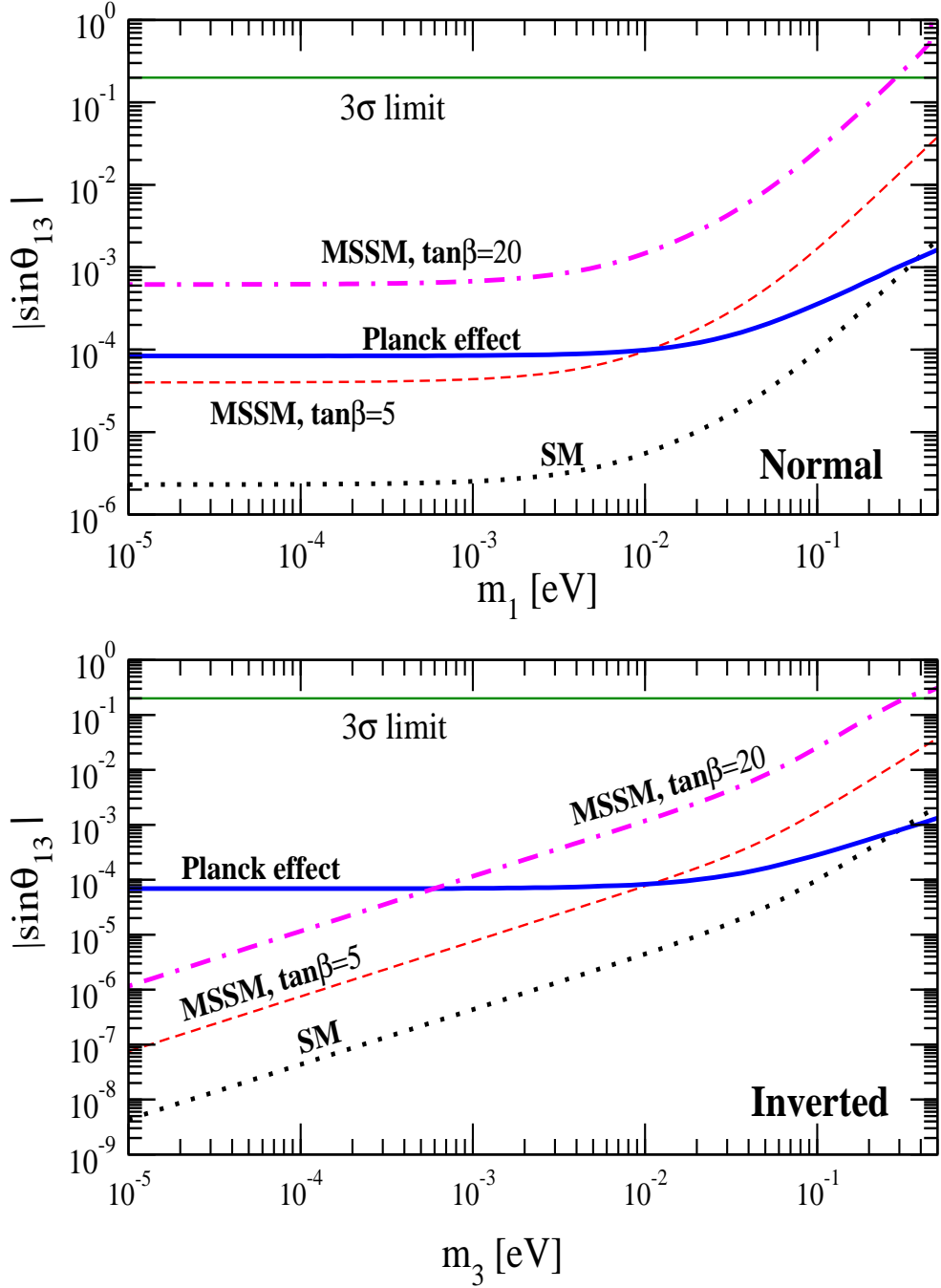} 
\caption{\label{theta13} Maximal generated value of the 
initially vanishing quantity $|\sin\theta_{13}|$ 
as a function of the smallest neutrino mass for the  
normal (upper panel) 
and inverted (lower panel) mass ordering.  
} 
\end{center} 
\end{figure}

\begin{figure}[hbt] 
\begin{center} 
\epsfig{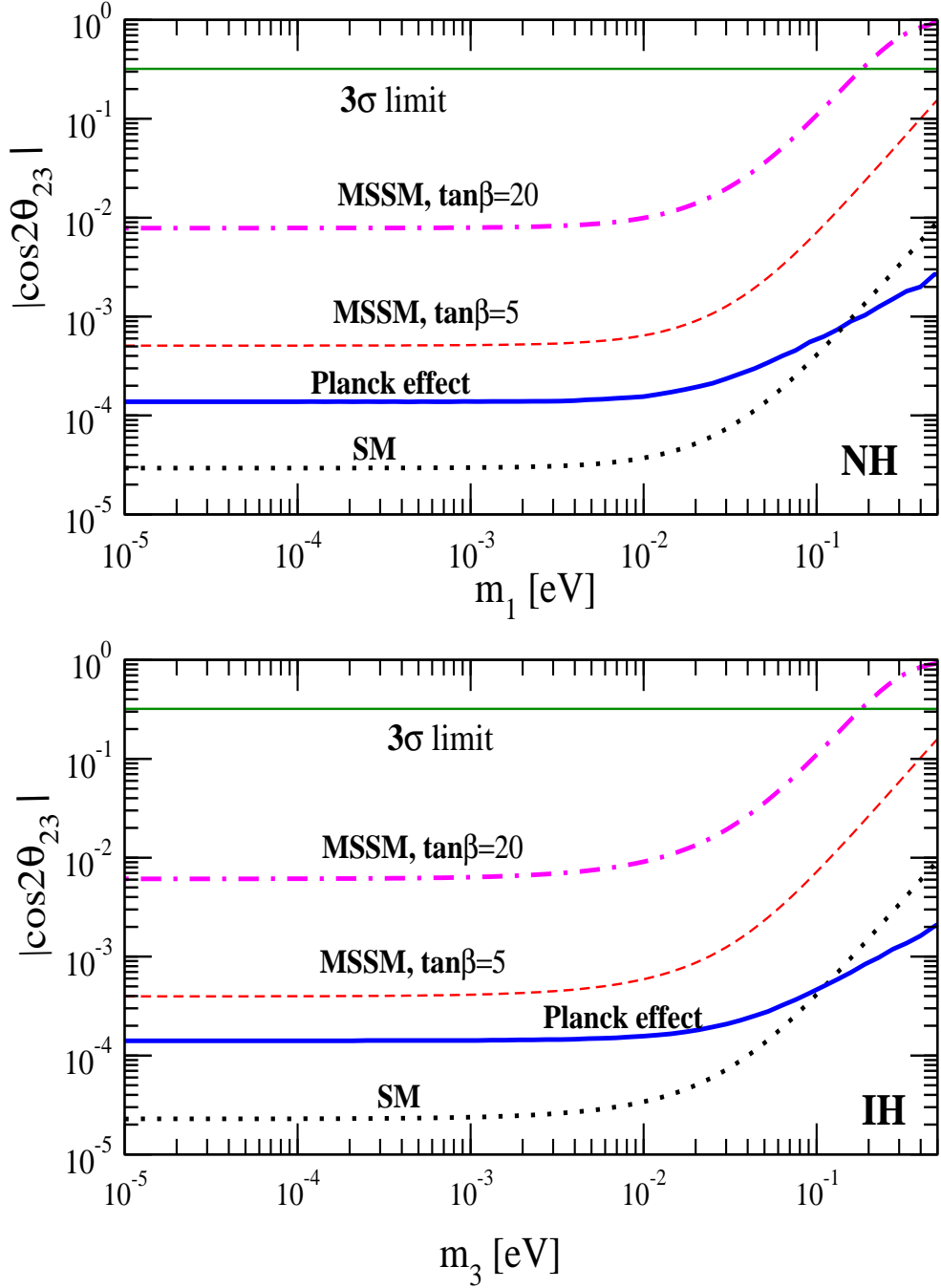} 
\caption{\label{theta23} Maximal generated value of the 
initially vanishing quantity $|\cos 2\theta_{23}|$ 
as a function of the smallest neutrino mass for the  
normal (upper panel) 
and inverted (lower panel) mass ordering. 
In the MSSM the sign of $\cos 2\theta_{23}$  
is negative (positive) for normal (inverted) 
ordering. The signs are reversed in the SM. 
} 
\end{center} 
\end{figure}

\begin{figure}[hbt] 
\begin{center} 
\epsfig{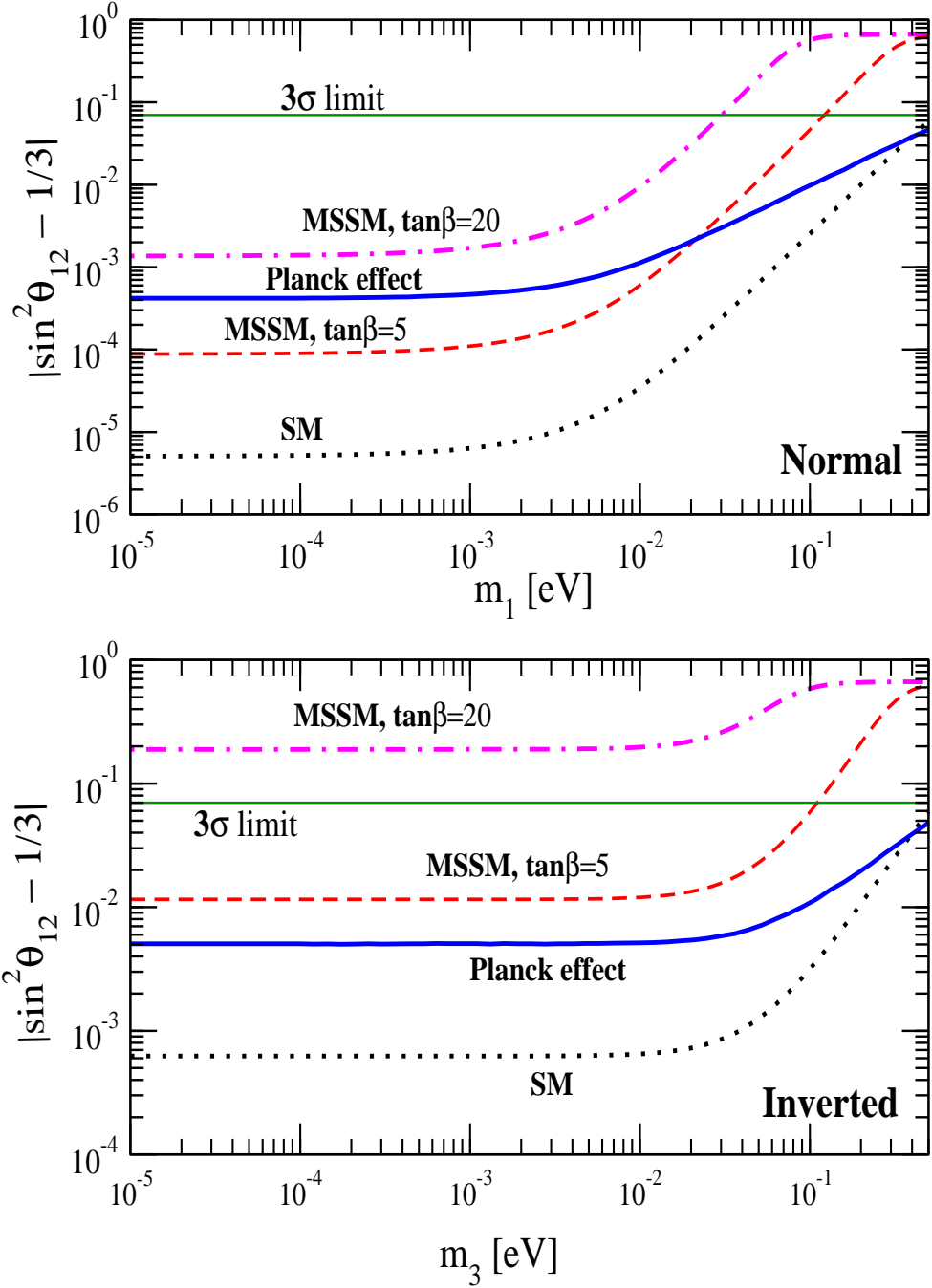} 
\caption{\label{theta12}Maximal generated value of the 
initially vanishing quantity $|\sin^2 \theta_{12} - 1/3|$ 
as a function of the smallest neutrino mass for the  
normal (upper panel) 
and inverted (lower panel) mass ordering. 
The sign of $(\sin^2 \theta_{12} - 1/3)$  
is always positive in the MSSM and always negative in  
the SM.  
} 
\end{center} 
\end{figure}

In Figs.~\ref{theta13}, \ref{theta23} and \ref{theta12} 
we show the maximal possible values of the  
initially vanishing quantities $|\sin \theta_{13}|$,  
$|\cos 2 \theta_{23}|$ and $|\sin^2 \theta_{12} - \frac 13|$  
generated from RG effects and from Planck scale effects,  
as a function  
of the smallest neutrino mass for normal and inverted  
mass ordering.  
The Figures for MSSM and SM 
are generated by numerically solving the RG equations in the small 
$\theta_{13}$ 
limit \cite{antusch-majorana}.  
 We have chosen in case of the MSSM  
$\tan \beta = 20$ and $\tan \beta = 5$.  
The relative magnitude of the deviations between these   
two cases should be $(1 + 5^2)/(1 + 20^2) \simeq 0.065$, which is  
confirmed by the plot. Also given in the plots are the maximal RG effects  
in the SM. 
As far as the mixing angles are concerned,  
for all practical purposes the  
results for SM can be obtained from the ones  
of MSSM with $\tan \beta = 20$ 
by multiplying them with $|(-3/2)/ (1 + 20^2)| \simeq  
0.0037$.  
We also indicate the present 3$\sigma$ bounds on the mixing parameters.

The RG running in SM is found to be very small. Even for 
values of $m_0$ in the QD regime the running stays within the 
present 3$\sigma$ limit for all three quantities.  
For MSSM the running is much stronger, especially for higher values  
of $\tan\beta$ and in the QD regime. 
For $| \sin^2 \theta_{12} - \frac 13| $ there is a plateau  
at $\sim$ 0.67 in the QD regime which corresponds to the maximum    
possible deviation of 2/3 in this quantity.  
As we are plotting the absolute values of the observables, we stress  
again that $\sin^2 \theta_{12} - \frac 13$ is always larger than  
zero in the MSSM.  
The same holds for $\cos 2 \theta_{23}$ in case of a inverted ordering,  
whereas for a normal ordering $\cos 2 \theta_{23}$ is negative.  
For the SM the signs are reversed. 
For the inverted ordering and small masses ($m_3 \ls 0.01$ eV), 
the deviation $\sin^2 \theta_{12} - \frac 13$ is roughly 
two orders of magnitude larger than that for the normal ordering.  
 
Note that since we show the maximal effects, the RG running beyond 
the 3$\sigma$ limits in the Figure 
does not imply that the corresponding 
neutrino mass values are ruled out. 
Rather, it implies that the Majorana phases can be constrained 
if the neutrino masses lie in the corresponding range.

The Figures also show the maximal possible values of the above 
three quantities  $|\sin \theta_{13}|$,  
$|\cos 2 \theta_{23}|$ and $|\sin^2 \theta_{12} - \frac 13|$  
due to Planck scale effects.  
They were obtained by numerically  
diagonalizing a mass matrix leading to TBM to which a flavor-blind  
Planck perturbation was added.  
It is to be noted that the values of  
$\cos 2 \theta_{23}$ and $\sin^2 \theta_{12} - \frac 13$ are symmetric  
about zero: depending on the values of the Majorana phases $\alpha_i$  
and the mismatch phases $\phi_i$, the 
Planck-corrected values of mixing angles can  
go in either direction. Thus these effects can either replenish or  
deplete the RG effects.  
The minimum values of the three quantities plotted in the Figure  
can be zero for suitable choices of the phases $\alpha_i$ and $\phi_i$.  
 
Let us compare both the effects: 
\begin{itemize} 
\item $\sin \theta_{13}$: the Planck effects can be larger than the RG effects  
in the SM unless the neutrino masses are above roughly 0.3 eV.  
Recall that the Planck corrections for QD neutrinos are proportional  
to $m_0$, whereas the RG corrections are proportional to $m_0^2$. 
For a normal ordering the Planck effects   
are always less than the maximal correction of $\tan \beta = 20$ but  
can exceed the corrections for $\tan \beta = 5$ if $m_1 \ls 0.01$ eV.  
For an inverted ordering Planck scale effects can exceed the  
corrections for $\tan \beta = 20$ (5) if $m_3 \ls 0.0005$ (0.01) eV;  
\item $|\cos 2 \theta_{23}|$: the Planck effects can be larger than the  
RG effects  
in the SM unless the neutrino masses are above roughly 0.1 eV. They  
are always less than the maximal correction of $\tan \beta = 20$  
and $\tan \beta = 5$;  
\item $|\sin^2 \theta_{12} - \frac 13|$:  
the Planck effects can be larger than the RG effects in the SM  
unless the neutrino masses are above roughly 0.5 eV. They  
are always below the maximal correction of $\tan \beta = 20$ and  
can exceed the corrections for $\tan \beta = 5$ if neutrinos  
are normally ordered and $m_1 \ls 0.01$ eV.  
\end{itemize}

\section{\label{sec:concl}Conclusions and Summary} 
 
We have studied the renormalization group and  
Planck scale corrections to neutrino mixing angles  
in the tri-bimaximal mixing scenario. Both these corrections 
need to be included while comparing the low energy neutrino 
mixing data with any postulated high scale mixing scenario. 
 
We give approximate expressions for the values of mixing angles  
at low scale starting from tri-bimaximal mixing at high scale for  
NH, IH and QD scenarios with RG running in the SM and the MSSM.  
We also plot the maximum RG effects as a function of the smallest  
neutrino mass for these scenarios.  
We find that in the SM the RG running has a negligible effect 
on the mixing angles. 
In the MSSM with large $\tan\beta$, while NH still gives 
 unobservably 
small deviations for all the mixing angles, 
IH is capable of generating significant running for $\theta_{12}$.  
In fact, matching $\theta_{12}$ with the data requires constraining 
the Majorana phases and $\tan \beta$ already at the present stage.  
For the QD scenario the running for all the three cases can be strong.  
The running depends crucially on the values of the Majorana phases and  
the neutrino mass scale: the corrections in the QD scenario 
grow as $m_0^2$.   
Precision measurements of neutrino mixing angles in future experiments  
should be able to put further constraints on the  
mass scale and Majorana phases if one assumes the TBM scenario.  
  
For the Planck scale effect we assume a flavor democratic 
dimension five operator at the Planck scale that contributes 
to the neutrino mass after electroweak symmetry breaking.  
We show that the corrections to the mixing angles  
can be quantified in terms of the ``mismatch phases'' $\phi_i$,
which are the values of the so-called ``unphysical'' phases
in the basis we have chosen. 
%depend in particular  
%on the phases $\phi_i$, which are usually called ``unphysical'' since they  
%have no impact in the absence of such Planck scale operators.  
Due to the special structure of the neutrino mass matrix giving rise  
to tri-bimaximal mixing, non-zero values of these phases are required  
for any Planck effect on the mixing angles and the masses $m_1$ and $m_3$.  
In general, if Planck scale effects are added to a $\mu$--$\tau$  
symmetric mass matrix then corrections to vanishing $U_{e3}$ and  
maximal $\theta_{23}$ are only possible if the unphysical  
phases have non-trivial values.

In the most general case when the mismatch
phases are non-vanishing, the Planck effects make 
the otherwise vanishing quantities $\sin\theta_{13}$, 
$\cos 2 \theta_{13}$ and $\sin^2 \theta_{12} - \frac 13$ grow  
almost linearly with the neutrino mass scale $m_0$ 
for quasi-degenerate neutrinos. 
The effects are in general largest for $\theta_{12}$.  
Even with a large value of ${\cal O}(\rm eV)$ for the neutrino masses,  
$\sin\theta_{13}$ and $\cos 2 \theta_{23}$ hardly exceed $10^{-3}$,  
and hence are virtually impossible to probe experimentally.  
However, deviations of $\sin^2 \theta_{12}$ from $1/3$ 
can be sizable, of the order of $0.1\,(m_0/{\rm eV})$. 
Deviations of a few degrees are thus allowed when neutrinos are  
quasi-degenerate. This deviation is larger than 
the resolution of the future precision $\theta_{12}$ experiments, 
and can be measured.

An interesting possibility, though hardly realizable in practice, 
 is the following: suppose one has measured deviations  
from tri-bimaximal mixing and knows the  
values of the Majorana phases and possibly of $\tan \beta$. In this case  
any additional correction beyond the RG effects  
to the mixing angles will stem from the Planck scale effects. As these  
depend on the mismatch phases, one could in principle  
extract some information on these phases. 
Moreover, if supersymmetry is not realized in nature, then  
the RG running is suppressed but the Planck scale effects can  
nevertheless inflict a sizable perturbation  
to the solar neutrino mixing angle, 
which can help us get a handle on these phases. 
 
In the case of the RG running, the signs of the correction to the  
mixing angles are predictable. For example in MSSM,  $\theta_{12}$ always 
increases from its high scale value, whereas $|\theta_{23}|$ 
increases (decreases) for normal (inverted) hierarchy. 
In the case of the Planck scale effects,  
the sign of $\cos 2 \theta_{23}$ and $\sin^2 \theta_{12} - 1/3$  
depends on all of the phases present, 
including the  
mismatch phases which will need a complete theory of
neutrino masses for their prediction. 
Therefore the Planck effects can either enhance or compensate the  
RG running. Constraining neutrino parameters due to running  
might therefore be not as straightforward as is usually done.  
If the neutrino mass is 0.5 (0.2) eV, then the modification  
from Planck scale effects to $\theta_{12}$ can be  
nearly $3^\circ$ $(1^\circ)$, which  
would weaken the constraints.  
Note that such a relaxation of constraints is applicable not only  
for TBM, but for any neutrino mixing scenario.

\section*{Acknowledgements} 
 
This work of W.R.~was supported by the ``Deutsche Forschungsgemeinschaft''  
under project number RO--2516/3--1 and partly by the Transregio  
Sonderforschungsbereich TR27 ``Neutrinos and Beyond''.  
The work of A.D.~is partly supported through the 
Partner Group project between the Max--Planck--Institut f\"ur  
Physik and the Tata Institute of Fundamental Research. 
S.G. acknowledges support from the Humboldt Foundation during the initial  
stage of this work.   
A.D. and S.G. would like to thank P. Roy for useful discussions.

\end{document}